\documentclass[pra,twocolumn,nofootinbib,preprintnumbers,superscriptaddress]{revtex4-1}

\usepackage{natbib}
\usepackage{amssymb,amsbsy,amsmath,amsfonts}
\usepackage{mathrsfs}
\usepackage{graphicx}
\usepackage{epsf,epsfig,float,latexsym,amsthm,fancyhdr,rotating}
\usepackage{graphics,psfrag,longtable}
\usepackage{slashed}
\usepackage{esint}
\usepackage{nicefrac}
\usepackage{braket}
\usepackage{mathtools}

\newcommand{\nn}{\nonumber}

\newcommand{\be}{\begin{equation}}
\newcommand{\ee}{\end{equation}}

\usepackage[colorlinks,citecolor=blue,linktoc=all,linkcolor=cyan]{hyperref} 

\DeclareMathAlphabet{\mathantt}{OML}{antt}{l}{it}
\DeclareMathAlphabet{\mathpzc}{OT1}{pzc}{m}{n}

\def\beq{\begin{equation}}
\def\eeq{\end{equation}}
\def\bea{\begin{eqnarray}}
\def\eea{\end{eqnarray}}
\def\beqa{\begin{equation}\begin{array}{l}}
\def\eeqa{\end{array}\end{equation}}

\def\barr{\left(\begin{array}{c}}
\def\earr{\end{array}\right)}
\def\bmat{\left(\begin{array}{cc}}
\def\emat{\end{array}\right)}

\begin{document}
\title {Polarization Transfer from the Twisted Light to an Atom}

\author{Andrei Afanasev}

\affiliation{Department of Physics,
The George Washington University, Washington, DC 20052, USA}

\author{Carl E. Carlson}

\affiliation{Physics Department, William and Mary, Williamsburg, Virginia 23187, USA}

\author{Hao Wang}

\affiliation{Department of Physics,
The George Washington University, Washington, DC 20052, USA}

\begin{abstract}
When polarized light is absorbed by an atom, the excited atomic system carries information about the initial polarization of light. For the light that carries an orbital angular momentum, or the twisted light, the polarization states are described by eight independent parameters, as opposed to three Stokes parameters for plane waves. We use a parameterization of the spin-density matrix of the twisted light in terms of vector and tensor polarization, in analogy with massive spin-1 particles, and derive formulae that define atom's response to specific polarization components of the twisted light. It is shown that for dipole ($S\to P$) atomic transitions, the atom's polarization is in one-to-one correspondence with polarization of the incident light; this relation is violated, however, for the transitions of higher multipolarity ($S\to D$, $S\to F$, {\it etc.})  We pay special attention to contributions of the longitudinal electric field into the matrix elements of atomic transitions. 
\end{abstract}
\date{\today
}
\maketitle


\section{Introduction}


Twisted photons \cite{Andrews-book,FrankeArnold_2017}, are a particular example of photons with a definite overall direction of propagation and with a shaped wave front that gives them unique properties.  In particular, twisted photons have an intrinsic orbital angular momentum in the direction of propagation whose value can be any integer value times $\hbar$.  This is an additional quantum number characterizing the photon state, which may allow applications in a number of areas, including quantum computing and communications, both encrypted and unencrypted.

A further feature of the twisted photon is its tensor polarization.  Plane wave photons, a spin one particle but with only two polarization states,  have no nontrivial tensor polarization.  Twisted photons have richer possibilities.  A given point of the wavefront of a twisted photon can be described in Fourier space as receiving contributions from momentum eigenstate photons with a variety of momenta, and the corresponding fields or vector potentials have components in all spatial directions, and can be usefully described in terms of basic $z$-direction (say) polarization vectors, specifically including a component in the longitudinal direction.  This gives the spin structure of the twisted photon state analogous  to the spin structure of a massive spin-1 state.  As a simple case, when the amplitudes of some state are $a_\pm$ and $a_0$ for polarization states in a spherical basis, the $T_{20}$ tensor polarization is
\be
T_{20} = \frac{1}{\sqrt{2}}\frac{ |a_+|^2 + |a_-|^2 - 2 |a_0|^2 }{  |a_+|^2 + |a_-|^2 + |a_0|^2 }  ,
\ee
clearly always $1/{\sqrt{2}}$ for a plane wave photon but not so for a twisted photon.  

It is well known that atomic collisions or absorption of light lead to polarization of atomic excited states that can be subsequently measured  ~\cite{Budker,Blum_2012}.  Early experimental work showing transfer of photon orbital angular momentum (OAM) to atoms can be found in~\cite{1999PhRvL..83.4967T,2003PhRvL..90m3001B}. Theoretical arguments for optical OAM transfer to {\it intrinsic} degrees of freedom of an atom were presented in Ref.\cite{Babiker2002}. The first experimental proof of new selection rules in the atomic photoexcitation by twisted photons was presented in Ref.\cite{2016NatCo...712998S} for atoms located on the optical vortex axis, followed by an off-axis analysis \cite{Afanasev_2018}. A comprehensive review of twisted-photon interactions with atoms can be found, $e.g.$, in Ref.\cite{FrankeArnold-PhilTran17}.

In this work we present a theoretical formalism that predicts OAM transfer and atomic polarization under absorption of the twisted photons.  


\section{Twisted Photons}

\subsection{Formalism}


We summarize some details regarding twisted photons that will prove useful in our discussion.  Much of the material included is known, and included for completeness and to make the paper self-contained.  However, we call attention to the analytic results quoted for amplitudes for individual photon polarization projections, which relates a collection of twisted photon amplitudes to a single plane-wave amplitude via Wigner functions and Clebsch-Gordan coefficients.   These results are new here albeit stylistically similar to earlier analytic results for the full amplitude~\cite{2014PhRvA..90a3425S}.  

A twisted photon propagating in the $z$-direction can be expanded in terms of plane wave states of helicity $\Lambda$ as
\be
\ket{ \kappa m_\gamma k_z \Lambda \, \vec \rho \, } = A \int \frac { d\phi_k }{ 2\pi } i^{-m_\gamma}
	e^{ i m_\gamma \phi_k }  e^{ -i \vec k \cdot \vec \rho}  \ket{ \vec k, \Lambda }		.
\ee
The $\ket{ \vec k, \Lambda }$ are plane wave states with momentum components  $\vec k = (k, \theta_k, \phi_k)$ in spherical coordinates, with all $k$ and $\theta_k$ the same, resulting in a Bessel beam, $c.f.$ \cite{PhysRevLett.58.1499}.  The factor $A$ is a normalization constant, given by $A = \sqrt{ \kappa/(2\pi) }$ in Serbo and Jentschura
 \cite{PhysRevLett.106.013001} with $\kappa = | \vec k_\perp | = k \sin\theta_k $ and $k_z = k \cos\theta_k$.  The variable $\vec \rho$ is a position vector with respect to the vortex center of the 
twisted photon in the $x$-$y$ plane.  A parameter $m_\gamma$ gives the full intrinsic angular momentum, projected in the $z$-direction, of the twisted photon.

The vector potential of the twisted photon in coordinate space is
\begin{align}
\label{eq:twistedwf}
\mathcal A^\mu_{\kappa m_\gamma k_z \Lambda}(x) &= -i \Lambda \, A \,
	e^{i(k_z z - \omega t + m_\gamma \phi_\rho )}	
				\nn\\	
	& \times \Bigg\{	e^{-i \Lambda \phi_\rho}  \cos^2\frac{\theta_k}{2} 	\,
	J_{m_\gamma-\Lambda}(\kappa\rho) \, \eta^\mu_\Lambda 
				\nn\\
	& \qquad + \frac{i}{\sqrt{2}}  \sin\theta_k	\,
	J_{m_\gamma}(\kappa\rho) \, \eta^\mu_0 
				\nonumber\\
&	\qquad 
	-   e^{ i \Lambda \phi_\rho}  \sin^2\frac{\theta_k}{2} 	\,
	J_{m_\gamma+\Lambda}(\kappa\rho) \, \eta^\mu_{-\Lambda}
	\Bigg\}	\,,
\end{align}
where the $\eta$'s are constant vectors,
\be
\eta^\mu_{\pm 1} = \frac{1}{\sqrt{2}}  \left( 0,\mp 1,-i,0 \right)	\,,
\quad \eta^\mu_0 =  \left( 0,0,0,1 \right)	\,.
\ee
This expression can be re-written in a compact form using Wigner matrices, 
\begin{align}
\label{eq:twistedwfcompact}
&\mathcal A^\mu_{\kappa m_\gamma k_z \Lambda}(x) = 
A\, e^{i(k_z z - \omega t)} \nn\\
&\times \sum_{\lambda=0,\pm 1} i^{-\lambda} d^1_{\lambda \Lambda}(\theta_k) 	J_{m_\gamma-\lambda}(\kappa\rho) \, \eta^\mu_\lambda e^{i( m_\gamma-\lambda) \phi_\rho}.
\end{align}

The amplitude for photoexciting an atomic state with principal and orbital quantum numbers 
$\{ n_f l_f m_f \}$ by a twisted photon is
\be
\mathcal M = \braket{  n_f l_f m_f | H_1 | n_i 0 0;  \kappa m_\gamma k_z \Lambda \, \vec b \, }	\,,
\ee
where $H_1 = (-e/m) \vec A \cdot \vec p$.   The initial state contains an S-state atom with principal quantum number $n_i$ and an incoming photon;  the final state has the atom in an excited state, and no photon. 

\begin{figure}[htbp]
\begin{center}
\includegraphics[width = 0.8 \columnwidth]{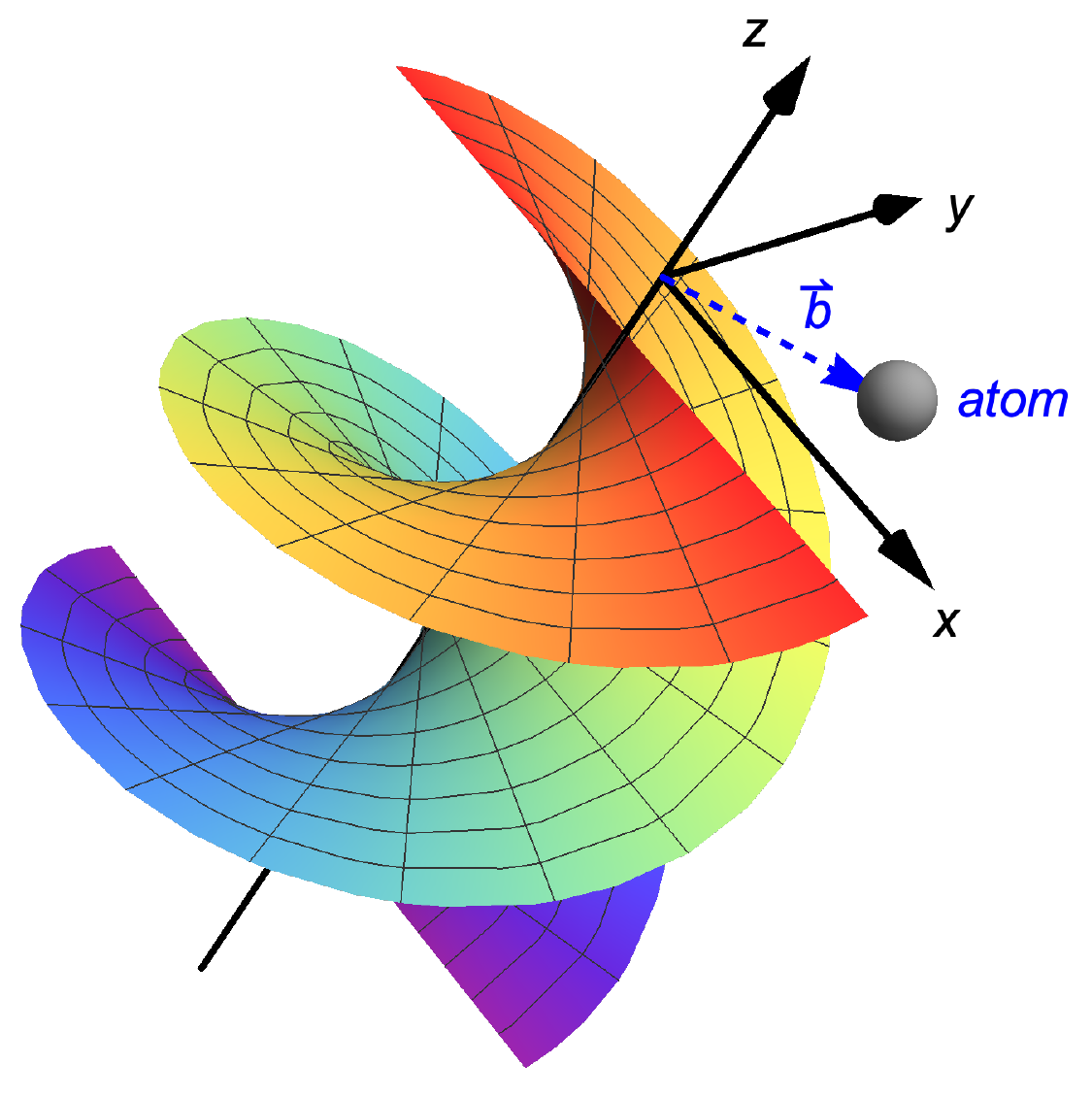}
\caption{The wavefront for a twisted photon. The helix is a surface of constant phase for the potential of Eq.~\eqref{eq:twistedwf} in the paraxial limit, for a given time and for $m_\gamma - \Lambda = -2$ in this example.  The target atom is also shown and is displaced from the photon's vortex line by impact parameter $\vec b$.}
\label{fig:atom-axes}
\end{center}
\end{figure}

The wavefront for a twisted photon, whose potential is given by Eq.~\eqref{eq:twistedwf}, is represented by the helix in Fig.~\ref{fig:atom-axes}. The figure shows for Eq.~\eqref{eq:twistedwf} a surface of constant phase in the paraxial limit (for a given time and for $m_\gamma - \Lambda = -2$).  The location of the target atom is also indicated, displaced from the photon's vortex line by impact parameter $\vec b$.


It is useful to recall that plane wave states propagating in the $z$ direction, $\vec k = k \hat z$, with helicity $\Lambda$ can be written in coordinate space as
\be
\eta_\Lambda  e^{i \vec k \cdot \vec r} = \sum_{L=0}^\infty i^L  \sqrt{ 4\pi ( 2L+1) } \, j_L(k r)
	Y_{L0}(\theta,\phi) \hat \eta_\Lambda  \,.
\ee
In Hilbert space, define states $\ket{ \vec k L M }$ so that the coordinate space representation is
\be
\braket{ \vec r \, | k \hat z, L M }  = i^L \sqrt{ 4\pi ( 2L+1 ) } j_L(kr) Y_{LM}(\theta,\phi)	\,,
\ee
whence
\be
\ket{ k \hat z, \Lambda } = \sum_{L=0}^\infty \ket{ k \hat z, L 0 } \ket{ \Lambda }	\,,
\ee
with photon polarization $\hat \eta_\Lambda$ represented by $\ket{ \Lambda }$.

Plane wave states in other directions are obtained by rotation.  Following Serbo and Jentschura ~\cite{PhysRevLett.106.013001} and the Wick 1962 convention~\cite{Wick:1962zz,Trueman:1964zzb},
\be
\ket{ \vec k, \Lambda } = R(\phi_k, \theta_k, 0) \ket{ k \hat z, \Lambda }
	= R_z(\phi_k) R_y( \theta_k ) \ket{ k \hat z, \Lambda }		\,.
\ee
Hence states moving in an arbitrary direction are given in terms of states moving in the $z$-direction as,
\begin{align}
\ket{ \vec k, \Lambda } &=  \sum_{L=0}^\infty  
		R_z(\phi_k) R_y( \theta_k ) \ket{ k \hat z, L 0 }	\ket{ \Lambda }		\nn\\
&= \sum_{L=0}^\infty \ \sum_{M = - L}^{M = L} \ \sum_{\lambda = -1}^1 
	e^{-i (M+\lambda) \phi_k}  
															\nn\\
& \quad	\times	\, d^{(L)}_{M0} (\theta_k)  \, d^{(1)}_{\lambda \Lambda} (\theta_k)  \,
			 \ket{ k \hat z, L M }	\ket{ \lambda }	
\end{align}

Substituting into the expression for the amplitude gives
\begin{align}
\mathcal M &= A	\int \frac { d\phi_k }{ 2\pi } \sum_{L,M,\lambda}  i^{-m_\gamma}
	e^{ i (m_\gamma - M -\lambda) \phi_k}  e^{ -i \kappa b \cos(\phi_k-\phi_b)}
						\nn\\
&	\quad  \times	\, d^{(L)}_{M0} (\theta_k)  \, d^{(1)}_{\lambda \Lambda} (\theta_k)  \,
			  \braket{  n_f l_f m_f | H_1 | n_i 0 0;  k \hat z, L M, \lambda }	\,.
\end{align}
Doing the $\phi_k$ integral leads to
\begin{align}
\mathcal M &= A	\sum_{L,M,\lambda}  	i^{ m_f - 2 m_\gamma }
	e^{ - i (m_f - m_\gamma) \phi_b}	J_{m_\gamma-m_f}( \kappa b )
											\\\nn
& \quad \times		d^{(L)}_{M0} (\theta_k)  \, d^{(1)}_{\lambda \Lambda} (\theta_k)  \,
			  \braket{  n_f l_f m_f | H_1 | n_i 0 0;  k \hat z, L M, \lambda }	\,.
\end{align}
The matrix element constrains $M = m_f - \lambda$.

Before evaluating the matrix element, consider the plane wave case.  
\begin{align}
&	\mathcal M^{(pw)} =  \braket{  n_f l_f m_f | H_1 | n_i 0 0;  k \hat z, \Lambda }
								\\\nn
&\!\! = \!	\sum_{L=0}^\infty 	\braket{  n_f l_f m_f | H_1 | n_i 0 0;  k \hat z, L 0, \Lambda }
  = \!\!  \sum_{L = l_f \pm 1}   \!\!  \mathcal M^{(pw)}_L	\, C^{l_f \Lambda}_{L0; 1 \Lambda}	,
\end{align}
where we have here defined what we might call the plane wave partial wave amplitude as 
\begin{align}
\mathcal M^{(pw)}_L &=   \frac{e}{m}   \, i^{L+1}  \frac{ 2L+1 }{ \sqrt{2l_f+1} }	\,  
	C^{l_f0}_{L0; 1 0}
						\nn\\
&	\quad \times		\int r^2 dr  \ R_{n_f l_f}(r) \,  j_L(kr) \, R'_{n_i 0}(r)  .
\end{align}

As a practical matter, for atomic applications the wavelength is long compared to the size of the region where the wave functions have support, so that $k = 1/ \lambdabar$ is small and the spherical Bessel functions ensure that only the lowest $L$ will contribute significantly.  Hence the full plane wave amplitudes is given in terms of just one of the plane wave partial wave amplitudes,
\begin{align}
\mathcal M^{(pw)} &=  \mathcal M^{(pw)}_{l_f-1}	\, C^{l_f \Lambda}_{l_f-1,0; 1 \Lambda}	.
\end{align}

To return to the matrix element needed for the twisted photon case, we can now find that it is given by the plane wave partial wave result with a differently-indexed Clebsch-Gordan coefficient, 
\begin{align}
\braket{  n_f l_f m_f | H_1 | n_i 0 0;  k \hat z, L M, \lambda } 
	= \mathcal M^{(pw)}_{l_f-1}	\, C^{l_f m_f}_{l_f-1,m_f-\lambda; 1 \lambda}  .
\end{align}
The full amplitude of atomic photoexcitation with a twisted photon becomes
\begin{align}
\label{eq:twistedMatEl}
\mathcal M &=  A \, i^{ m_f - 2 m_\gamma }
	e^{ - i (m_f - m_\gamma) \phi_b}	J_{m_\gamma-m_f}( \kappa b ) \,
	\mathcal M^{(pw)}_{l_f-1}
							\nn\\
&\quad	\times	\sum_{\lambda}	\,
	d^{(l_f-1)}_{m_f-\lambda, 0} (\theta_k)  \, d^{(1)}_{\lambda \Lambda} (\theta_k)	\,
		C^{l_f m_f}_{l_f-1,\,m_f-\lambda; 1 \lambda}	\,.
\end{align}

Finally, the atomic form factors $g$ we previously defined in~\cite{Afanasev:2013kaa} are the coefficients of the 
$d^{(1)}_{\lambda\Lambda}(\theta_k)$ Wigner functions, or
\begin{align}
\label{eq:gfactors}
g_{n_f l_f m_f \lambda} &=   A \, i^{ m_f - 2 m_\gamma }
	e^{ - i (m_f - m_\gamma) \phi_b}	J_{m_\gamma-m_f}( \kappa b ) \,	\mathcal M^{(pw)}_{l_f-1}
												\nn\\	
&	\quad	\times	\ 	d^{(l_f-1)}_{m_f-\lambda, 0} (\theta_k)  	\,	
			C^{l_f m_f}_{l_f-1,\,m_f-\lambda; 1 \lambda}		\,,
\end{align}
for a given initial state defined by $n_i$, $m_\gamma$, and $\Lambda$. 

Summation in Eq.(\ref{eq:twistedMatEl}) over the index $\lambda$ yields an expression

\begin{align}
							\nn\\
&\quad	 \mathcal M^{(pw)}_{l_f-1} \sum_{\lambda}	\,
	d^{(l_f-1)}_{m_f-\lambda, 0} (\theta_k)  \, d^{(1)}_{\lambda \Lambda} (\theta_k)	\,
		C^{l_f m_f}_{l_f-1,\,m_f-\lambda; 1 \lambda}  \nn\\
		&=\mathcal M^{(pw)}_{l_f-1} 	C^{l_f m_f}_{l_f-1,m_f-\lambda; 1 \Lambda}d^{(l_f)}_{m_f \Lambda} (\theta_k) \\
		&=\mathcal M^{(pw)} d^{(l_f)}_{m_f \Lambda} (\theta_k), \nn
\end{align}
thereby recovering Eq.(19) of Ref.\cite{2014PhRvA..90a3425S}. The result of Eq.(\ref{eq:twistedMatEl}) enables us to analyze atomic response to specific components of the electric field of the incident twisted light represented by the index $\lambda$.

The expressions for atomic form factors $g$  dramatically simplify for the case of dipole $S\to P$ transitions. Indeed, setting $l_f=1$ in Eq.(\ref{eq:gfactors}), we obtain
\begin{align}
&g_{n_f l_f m_f \lambda} = \nn\\    &A \, i^{ m_f - 2 m_\gamma }
	e^{ - i (m_f - m_\gamma) \phi_b}	J_{m_\gamma-m_f}( \kappa b ) \,	\mathcal M^{(pw)} \delta_{m_f \lambda},
\end{align}
which shows one-to-one mapping of the electric field spherical components $\lambda$ onto the magnetic quantum numbers $m_f$ of the excited atom,
relating the $E1$ dipole transition matrix element for a given final magnetic quantum number $m_f$ to the vector potential component $\lambda$ as follows:
\begin{equation}
 \mathcal M=-\eta_{\mu; \lambda} A^\mu_{\kappa m_\gamma k_z \Lambda}(x)|_{z,t=0}\mathcal M^{(pw)} (-1)^{(m_f-m_\gamma)}\delta_{\lambda m_f}.
\end{equation}
It immediately leads to the relation between the spin density matrices of the twisted photon $\rho^{(\gamma)}_{\lambda \lambda'}$ and the excited atom $\rho^{(e)}_{m_f m_f'}$,
\begin{equation}
\label{eq:sdm}
\rho^{(\gamma)}_{\lambda \lambda'}=\rho^{(e)}_{m_f m_f'} (-1)^{(m_f-m_f')} \delta_{\lambda m_f} \delta_{\lambda' m_f'},
\end{equation}
where we adopt the definition for the spin-density matrix of a twisted photon in analogy with a massive spin-1 particle. The relation Eq.(\ref{eq:sdm}) provides a tool to measure twisted photon's spin density matrix by measuring polarization states of the excited atom (through polarization and angular distribution of radiation from its de-excitation). It also unambiguously predicts spin polarization of the excited atom if the twisted photon's spin-density matrix is known.

We have only considered electric transitions, which do not affect the spin of the target electrons.  One can also include magnetic transitions, in which case an additional set of plane wave amplitudes can contribute.  In special circumstances, interesting results occur when there are two different amplitudes of comparable magnitude.  Such results have been discussed in~\cite{PhysRevA.97.023422,PhysRevA.100.043416}, albeit without the emphasis on separating the longitudinal and traverse contributions that we are providing here.


\subsection{An application of the analytic atomic form factors}


A feature of the twisted photon wavefront is the presence of longitudinal electromagnetic fields.  These fields are often neglected in a paraxial approximation, but should not be neglected in general.  They can have dramatic effects on transition amplitudes in certain cases.  A example was given by Quintiero et al.~\cite{2017PhRvL.119y3203Q}.  We can study this example here to see how the analytic expressions for the $g$-factors, $g_{n_f l_f m_f \lambda}$, quickly show the consequences of the longitudinal fields.

The Quintiero et al.~\cite{2017PhRvL.119y3203Q} example studies $m_\gamma = 0$ twisted photons.  These have, in relative size, particularly large longitudinal fields in their makeup.  The specific example was for this $m_\gamma$ with negative helicity, $\Lambda = -1$.   For comparison they also calculate photoexcitation by a $m_\gamma = 2, \Lambda  = +1$ state.  The initial atomic state in each case is the $4S_{1/2}$ state of $^{40}$Ca, with electron polarization in the negative direction, and the final states are $3D_{5/2}$ with the total angular momentum projection $m_j$ being $-1/2$ and $3/2$ for the two cases.

Given our results above, the matrix elements are given in terms of an overall factor and a collection of $d$-functions and Clebsch-Gordan factors,
\begin{align}
&\mathcal M(m_\gamma=0, m_j = -1/2) = \sqrt{\frac{3}{5}} A \,\mathcal M^\text{pw}_1  J_0(\kappa b) \times
		\nn\\
&	\Big\{
	d^{(1)}_{1,-1}(\theta_k)	d^{(1)}_{-1,0}(\theta_k)	C^{20}_{1,-1;11}	
		+  d^{(1)}_{0,-1}(\theta_k)	d^{(1)}_{0,0}(\theta_k)	C^{20}_{10;10}	  \nn\\
&\quad	+   d^{(1)}_{-1,-1}(\theta_k)	d^{(1)}_{1,0}(\theta_k)	C^{20}_{11;1,-1}	\,
\Big\}															\nn\\
&= -\sqrt{\frac{3}{5}} A \,\mathcal M^\text{pw}_1	J_0(\kappa b) \frac{ \sin\theta_k }{ 2\sqrt{3} }
\left( - 3 \cos\theta_k	
\right)	.
\end{align}
The first factor is an additional Clebsch-Gordan coefficient that projects the orbital $\{l_f,m_f\} = \{2,0\}$ state that we calculate onto the total angular momentum $\{j,m_j\} = \{5/2,-1/2\}$ state the experimenters measure; for atomic excitation with the twisted photons, it was introduced in Ref.\cite{Afanasev_2018} . Of most interest are the middle terms, which come from the longitudinal fields.  

Similarly, the comparison matrix element is
\begin{align}
&\mathcal M(m_\gamma=2, m_j = 3/2) 		\nn\\
&	\qquad\qquad	= - \sqrt{\frac{1}{5}} A \,\mathcal M^\text{pw}_1 J_0(\kappa b) 
	d^{(1)}_{1,1}(\theta_k)	d^{(1)}_{1,0}(\theta_k)	C^{22}_{11;11}															\nn\\
&	\qquad\qquad	= + \sqrt{\frac{1}{5}} A \,\mathcal M^\text{pw}_1	J_0(\kappa b) 
		\frac{ \sin\theta_k }{ \sqrt{2} }
\cos^2\frac{\theta_k}{2}	,
\end{align}
other terms being precluded by quantum numbers.

One can take the ratio, and then give the result in the paraxial (small $\theta_k$) limit,
\begin{align}
\left| 
\frac{  \mathcal M(m_\gamma=0, m_j = -1/2)  }{  \mathcal M(m_\gamma=2, m_j = 3/2)  } \right| = 
\begin{cases}
\displaystyle{  \frac{3}{ \sqrt{2} }  } = 2.123	&	\text{if } A_z \ne 0	\\[2.2ex]
\displaystyle{  \frac{1}{ \sqrt{2} }	  } = 0.707	&	\text{if } A_z = 0		\,,
\end{cases}
\end{align}
in agreement with~\cite{2017PhRvL.119y3203Q}, with the first also in agreement with the experimental result~\cite{2016NatCo...712998S} of $2.21 \pm 0.13$. 
Thus, the longitudinal electric field is responsible for 2/3 of the magnitude for $S\to D$, $E2$-transition amplitude with $\Delta m=0$. Whereas the results of Ref.~\cite{2017PhRvL.119y3203Q} were obtained for an atom placed the beam center $b=0$, the above formulae are valid away from the beam center, as well.

The equation (\ref{eq:twistedMatEl}) predicts that the longitudinal electric fields only contribute to $\Delta m=0$ transitions for $E1$ atomic excitations, but the situation changes for $E2$ and higher multipoles. In the latter case longitudinal fields also contribute into $\Delta m\neq0$ transitions, while $\Delta m=0$ ones are partially driven by transverse fields, as well.


 \subsection{Spin Density Matrices for the Twisted Photons and Excited Atoms}
 
 
 In photoexcitation by twisted light, the twisted photons lead to polarization of the final atomic system that is very different from the results from plane wave photoexcitation.  One can characterize the polarization of the final atomic states by means of the vector, tensor, an higher polarizations.  In this section we shall list some of the definitions associated with these polarizations, and call attention to a remarkable relation between the photon and atomic polarizations that is valid for $S\to P$ wave transitions.  The following section will show some example that give remarkable results.   As a reminder, we are in this paper concentrating on electric transitions, which do not affect the atomic spin and hence will speak mainly about results in terms of the orbital atomic wave functions. 
 
 For the general case of transition to angular momentum $\ell$,  when the final state density matrix is $\rho_{m'_f m_f}$ (in a spherical basis with quantization axis in the direction of propagation of the photon), the general expression for the tensor polarizations is~\cite{Ohlsen_1972,varshalovich,Blum_2012}
 \begin{equation}
T_{KM}(\ell)=\sqrt{2K +1} \sum_{m_f,m_f'}  \rho_{m'_f m_f}  C^{\ell m_f'}_{\ell m_f; KM},
\end{equation}
where the $C^{\ell m_f'}_{\ell m_f; KM}$ are Clebsch-Gordan coefficients.  We follow the normalization from~\cite{Blum_2012}; the expressions in~\cite{Ohlsen_1972,varshalovich} are the same but with an additional factor $1/\sqrt{2\ell +1}$.  

We can specialize to $T_{K0}(\ell)$, where~\cite{Blum_2012} also has the alternative notation of orientation (or alignment) parameters $B_K(\ell) = T_{K0}(\ell)$, with 
\begin{equation}
B_K(\ell)=(2\ell +1)^\frac{1}{2}\sum_{m_f}(-1)^{J-m_f}C^{K0}_{\ell m_f;\ell, -m_f} \, w(m_f),
\end{equation}
recalling that the diagonal elements of the density matrix are equal to the probabilities $w(m_f)$ of finding the atomic state with projection $m_f$.

Explicitly, the expressions for these parameters read for spin-1,
\begin{align}
B_1(1)&= T_{10}(1) = \sqrt{\frac{3}{2}}(w(1)-w(-1)) \nn\\
B_2(1)&= T_{20}(1) = \sqrt{\frac{1}{2}}(w(1)-2w(0)+w(-1)),
\end{align}
and for spin-2,
\begin{align}
B_1(2)&=\sqrt{\frac{1}{2}}(2w(2)+w(1)-w(-1)-2w(-2)) \\
B_2(2)&=\sqrt{\frac{5}{14}}(2w(2)+2w(-2)-w(1)-w(-1)-2w(0)) \nn\\
B_3(2)&=\sqrt{\frac{1}{2}}(w(2)-2w(1)+2w(-1)-w(-2)) \nn\\
B_4(2)&=\sqrt{\frac{1}{14}}(w(2)-4w(1)+6w(0)-4w(-1)+w(-2)). \nn
\end{align}
One may recognize that the parameter $B_1$ is proportional to the expectation value $\langle \ell_z \rangle$ of atom's angular momentum projection,
\begin{equation}
\label{eq:Lz}
\langle \ell_z \rangle =\sum_{-\ell<m_f<\ell} m_f w(m_f).
\end{equation}
Also, in general, the upper and lower limits on the $B_K$ are asymmetric, for example
\be
- \sqrt{2} \le B_2(1) = T_{20}(1) \le \frac{1}{\sqrt{2}} .
\ee

One can already see possibilities for differences from photoexcitation with plane wave photons, where the angular momentum in the propagation direction comes only from the spin, and the only possible final states have $m_f = \Lambda = \pm 1$.  This means that $T_{K0}$ is a fixed number for plane wave photons, and as examples, one finds $T_{20}(1) = 1/ \sqrt{2}$ and $T_{20}(2) = - \sqrt{5/14}$, and this is true regardless of the location of the target atom.

Becoming more specific and concentrating on $S \to P$ transitions caused by twisted photons, there is a remarkable feed-through of the full set of tensor polarizations between the photon and the atomic final state.  We will find it convenient to use Cartesian components in what follows.

An arbitrary (pure) spin state of spin-1 particle $|\chi_1 \rangle$, or a $p$-state of an atom, can be expanded in terms of basis kets $|\chi_{1i}\rangle$ $(i=x,y,z)$ in a Cartesian basis, 
$|\chi_1\rangle= \sum_i a_i |\chi_{1i}\rangle$, as well as in a spherical basis $|\chi_1 \rangle=a^+|\chi_{1+} \rangle+a^0|\chi_{10} \rangle+a^-|\chi_{1-} \rangle$,  where $a_i$ and $a^{\pm}$ are complex amplitudes related by  $$ a^\pm=\frac{(\mp a_x+i a_y)}{\sqrt{2}}, \, a_0=a_z, \, \sum_i a_i a_i^*=1.$$
  
The spin density matrix is now $\rho_{ij}=a_i a_j^*$, and we will use parameterization and normalization conventions from Ref.\cite{Ohlsen_1972}:
\begin{align}
\rho=&\frac{1}{3}\Big\{ I+\frac{3}{2}(p_x \mathcal P_x+p_y \mathcal P_y+p_z \mathcal P_z)+ \nn\\
&\frac{2}{3}(p_{xy} \mathcal P_{xy}+p_{yz} \mathcal P_{yz}+p_{xz} \mathcal P_{xz})+\\
&\frac{1}{6}(p_{xx}-p_{yy})(\mathcal P_{xx}-\mathcal P_{yy})+\frac{1}{2}p_{zz}\mathcal P_{zz}), \nn
\end{align}
where $\mathcal P_i,\, \mathcal P_{ij}$ are the operators of spin and quadrupole moment,  and $p_i,\, p_{ij}$ are corresponding vector and quadrupole polarizations that can be expressed in terms of the above amplitudes $a_i$ and $a^{\pm,0}$,
\begin{equation}
p_i=i \sum_{jk} \epsilon_{ijk}a_i a_k^*; \, p_{ik}=-\frac{3}{2} (a_i a_k^*+a_k a_j^*-\frac{2}{3}\delta_{ik}).
\end{equation}
In particular,
\begin{align}
&p_{xx}-p_{yy}=3 (a^+a^{-*}+a^-a^{+*});\nn\\ &p_{zz}=|a^+|^2+|a^-|^2-2|a^0|^2 ; \nn\\
&p_{xy}=i\frac{3}{2}(a^+ a^{-*}-a^-a^{+*}) \\
&p_{xz}=\frac{3}{2\sqrt{2}}(a^+a^{0*}+a^0a^{+*}-a^-a^{0*}-a^{0}a^{-*}) \nn \\
&p_{yz}=i\frac{3}{2\sqrt{2}}(a^+a^{0*}-a^0a^{+*}+a^-a^{0*}-a^{0}a^{-*}). \nn
\end{align}
The components of polarization are bound as follows, $-3\leq p_{xx}-p_{yy} \leq 3$, $-2\leq p_{ii} \leq 1$, $-\frac{3}{2}\leq p_{ij}\leq \frac{3}{2}$ ($i\neq j$), and $-1\leq p_i\leq1$.

For $S\to P$ atomic excitation with the twisted photons, the amplitudes $a^{\pm,0}$ are replaced by either corresponding $\lambda=\pm,0$-component of the electric field (\ref{eq:twistedwf}) or the transition amplitude (\ref{eq:twistedMatEl}) for a final state with a given magnetic quantum number $m_f=\pm,0$. It follows from Eq.(\ref{eq:sdm}) that in this case the following relations hold between spin density matrices of the twisted photons and the excited $\ell = l_f=1$ atomic state:

\begin{align}
\label{eq:pols2p}
&p^{(\gamma)}_{xx}-p^{(\gamma)}_{yy}=p^{(e)}_{xx}-p^{(e)}_{yy},\nn\\ 
&p^{(\gamma)}_{zz}= p^{(e)}_{zz}, \, p^{(\gamma)}_{xy}= p^{(e)}_{xy}\\
&p^{(\gamma)}_{xz}=-p^{(e)}_{xz}, \, p^{(\gamma)}_{yz}=-p^{(e)}_{yz}\nn\\
&p^{(\gamma)}_{z}=p^{(e)}_{z}, \, p^{(\gamma)}_{x,y}=-p^{(e)}_{x,y}\nn
\end{align}


\subsection{Numerical Examples of Atomic Polarization by the Twisted Photons}


Given the definitions in the previous section, and the calculations for the twisted photon photoexcitation amplitudes as in Eq.~\eqref{eq:twistedMatEl}, we can obtain the density matrix for a given set of final states, or the probabilities of finding given magnetic quantum numbers, and then give numerical and graphical results for the alignment parameters and tensor polarizations for selected examples.  We choose the angle parameter $\theta_k$=0.1 rad in these examples, in order to match the experimental setup with  trapped $^{40}$Ca ions of Ref.\cite{2016NatCo...712998S,Afanasev_2018}.

\begin{figure*}[h]
\centering
\includegraphics[scale = 0.4]{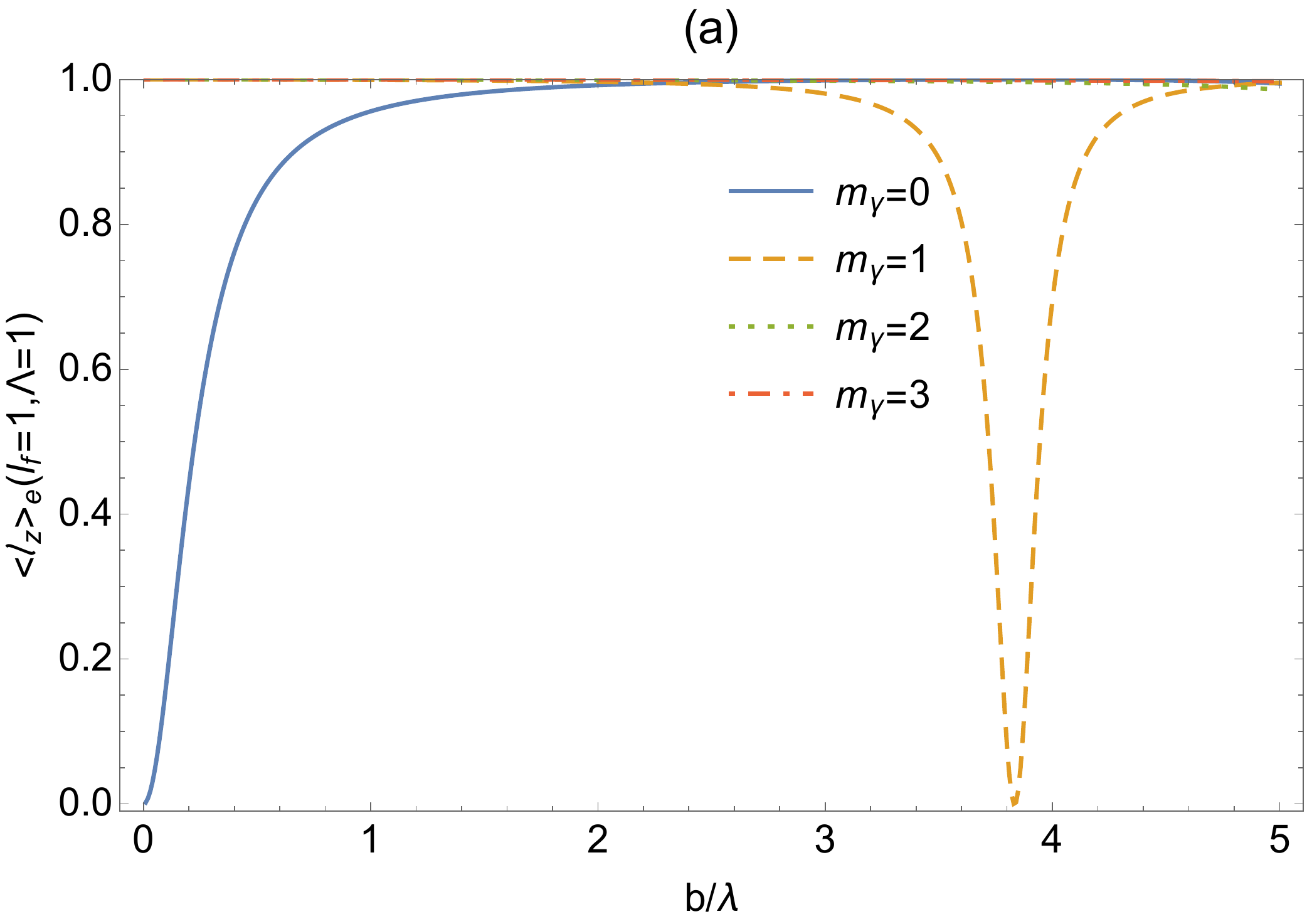} \hfil
\includegraphics[scale = 0.4]{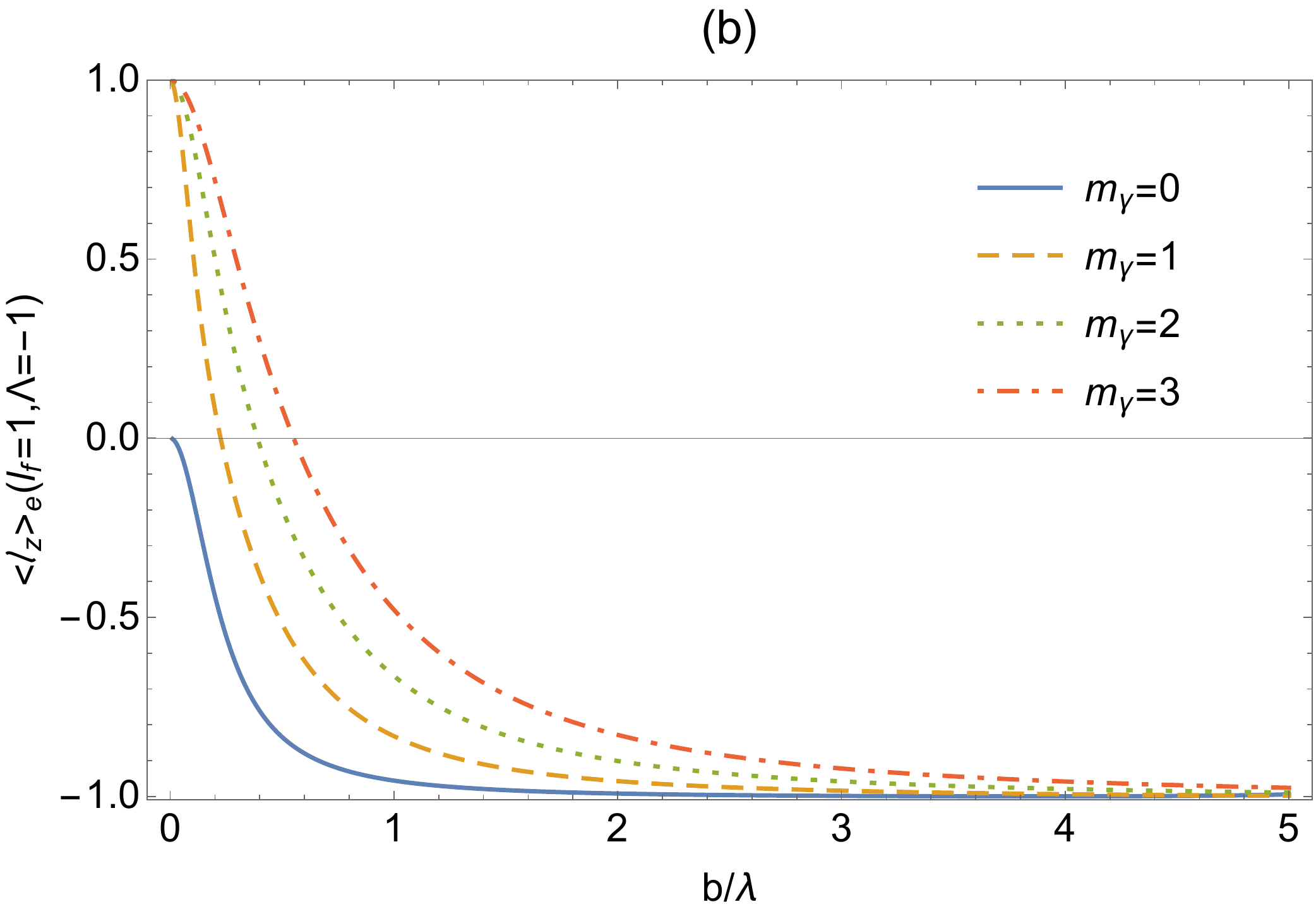}

\includegraphics[scale = 0.4]{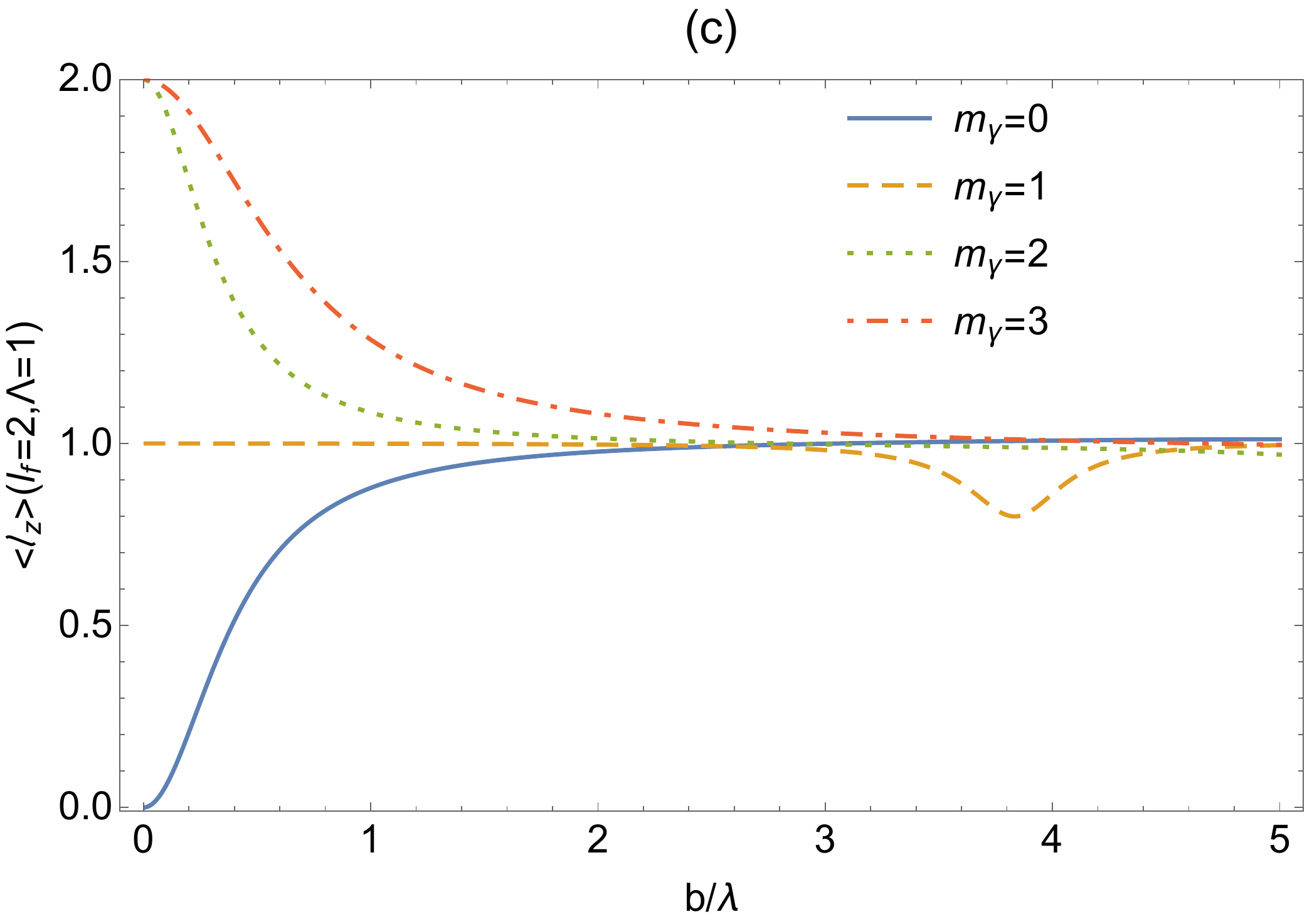} \hfil
\includegraphics[scale = 0.4]{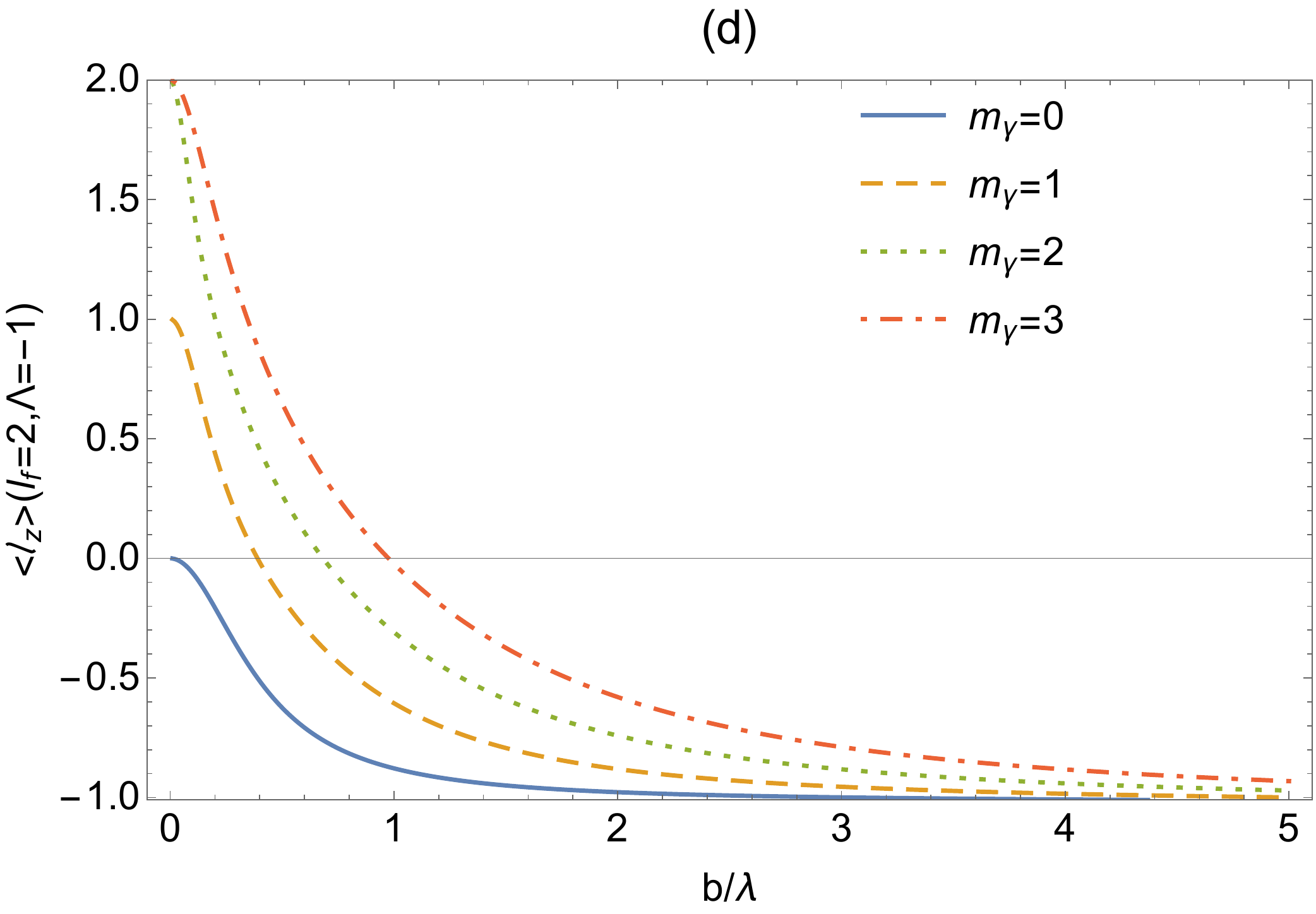}
\caption{Mean angular momentum transfer Eq.(\ref{eq:Lz}) along the beam direction passed by the twisted light to atom's internal degrees of freedom in an $S\to P$ transition (a,b) and $S\to D$ transition (c,d). For both pairs, $\Lambda = +1$ is on the left and $\Lambda =-1$ is on the right.  The horizontal axis shows atom's position $b$ with respect to the vortex center measured in units of light's wavelength. See text for notations.}
\label{fig:Lz}
\end{figure*}

We begin by showing in Fig.~\ref{fig:Lz} the average angular momentum projection, $\braket{l_z}$, Eq.~\eqref{eq:Lz}, transferred to the atomic state as a function of impact parameter, for the examples of $S \to P$ and $S \to D$ transitions for several choices of incoming twisted photon quantum numbers.  The $m_\gamma$ and $\Lambda$ chosen are labeled on the Figure.  We see that at low impact parameter $b$, there are signifiant deviations from what would be expected from a plane wave with corresponding helicity, and for the $S\to D$ case, there are angular momentum transfers in excess of one unit of $\hbar$.   For larger impact parameters, one relaxes to a tendency, though not an invariable one, to match the plane wave expectation. The plots of Fig.~\ref{fig:Lz} agree with the original observation of Babiker and collaborators Ref.\cite{Babiker2002}: One needs E2 transitions and higher to pass twisted light's angular momentum $>\hbar$ to (internal) atomic excitation. The magnitude of $\braket{l_z}$ in Fig.~\ref{fig:Lz} justifies "superkick" results by Berry and Barnett \cite{Barnett_2013} who predicted large atomic recoil in single-photon absorption in the vicinity of phase singularity, since OAM of recoil equals the difference $m_{\gamma}-\braket{l_z}$ \cite{Afanasev:14}.

Next we turn to tensor polarization.  Interesting examples come by making the initial photon a superposition of states with different $m_\gamma$ and $\Lambda$.  Such superpositions are experimentally feasible and have been produced and used for example in~\cite{2019arXiv190207988S}.    

Here we specifically choose an initial state that is a coherent superposition of states with ($m_\gamma=-2$, $\Lambda=1$) and ($m_\gamma=3$, $\Lambda=-1$).

\begin{figure*}[hb]
\centering
\includegraphics[scale = 0.43]{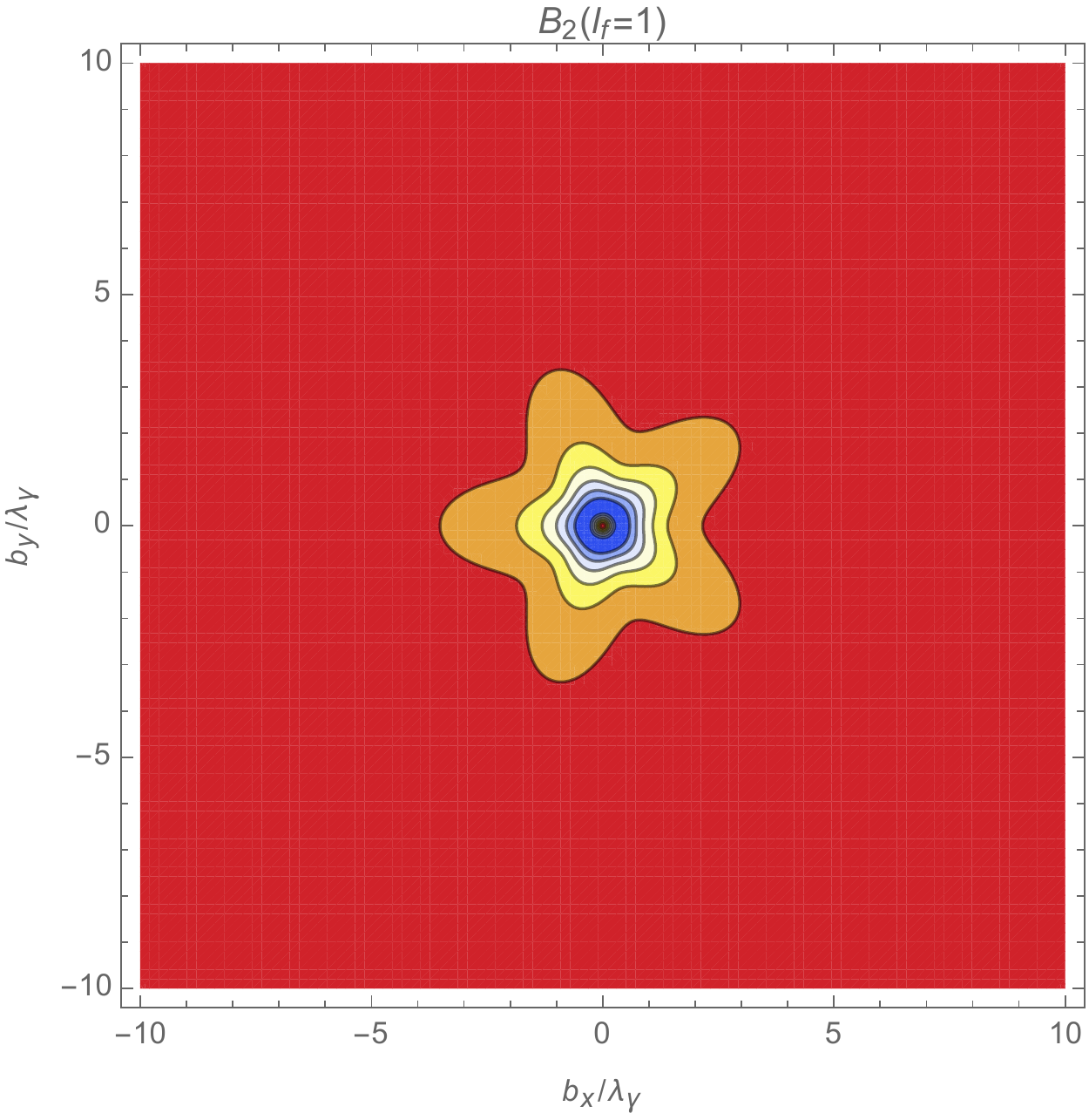} \hfil
\includegraphics[scale = 0.48]{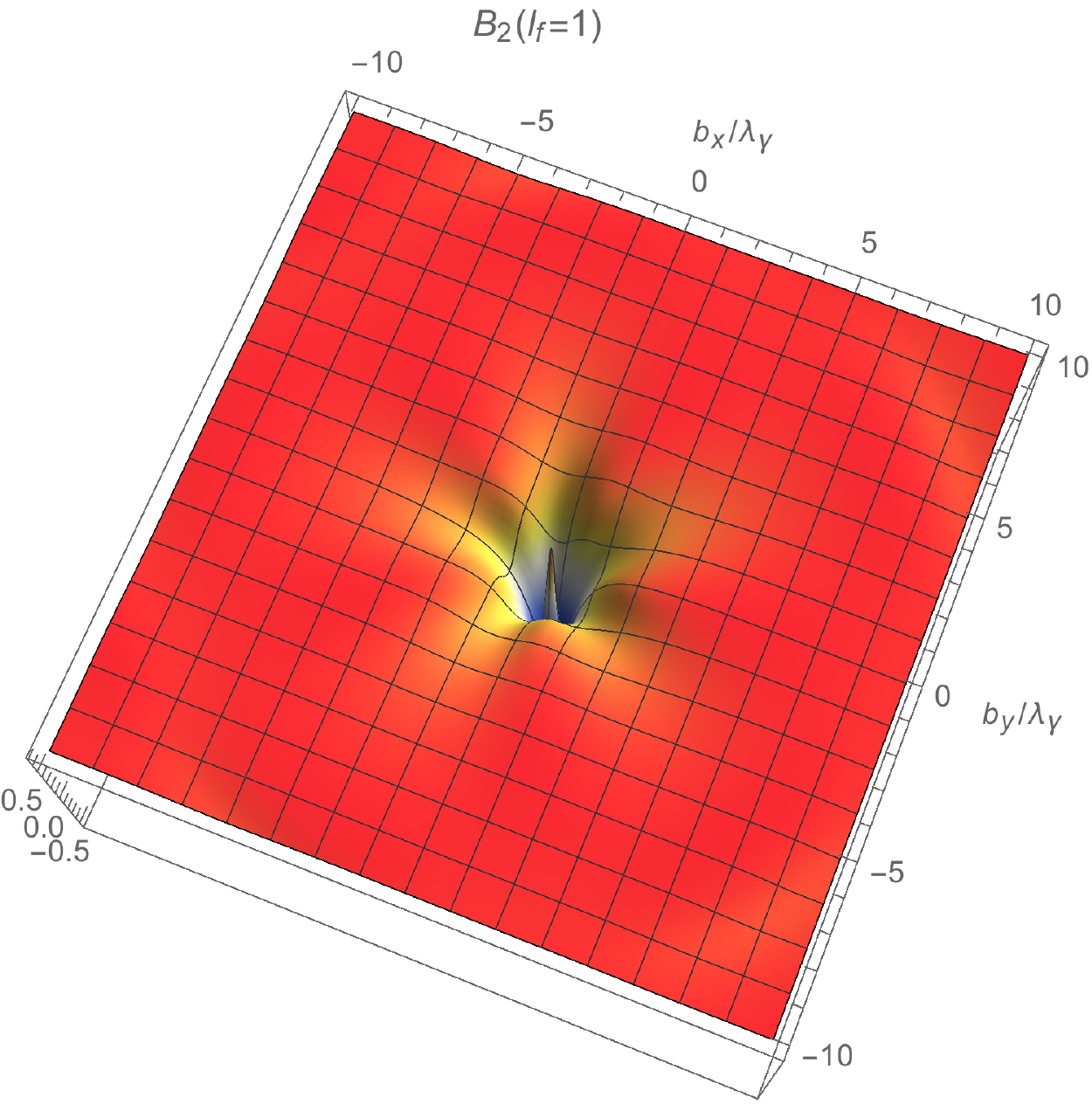}

\includegraphics[scale = 0.43]{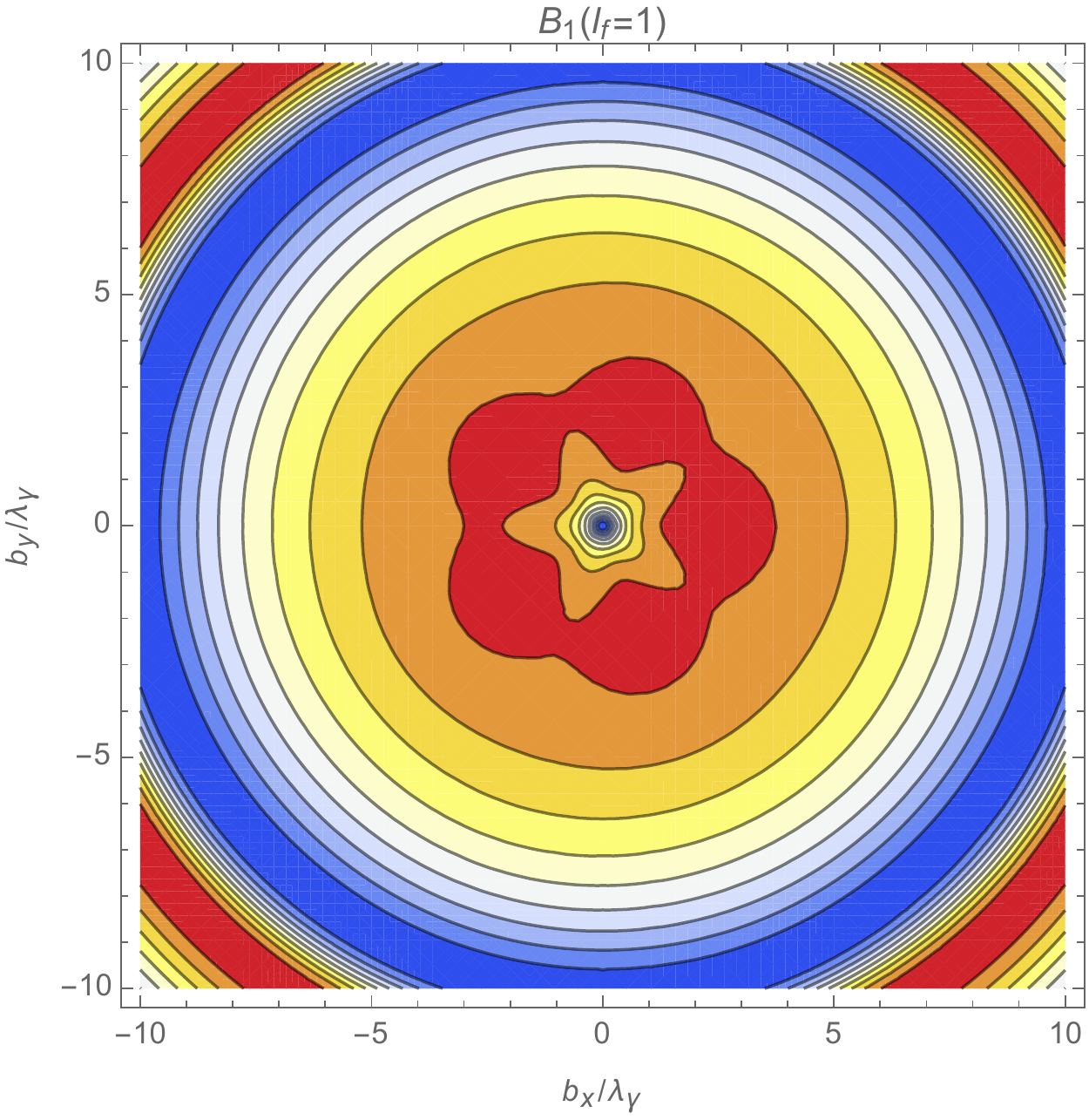} \hfil
\includegraphics[scale = 0.48]{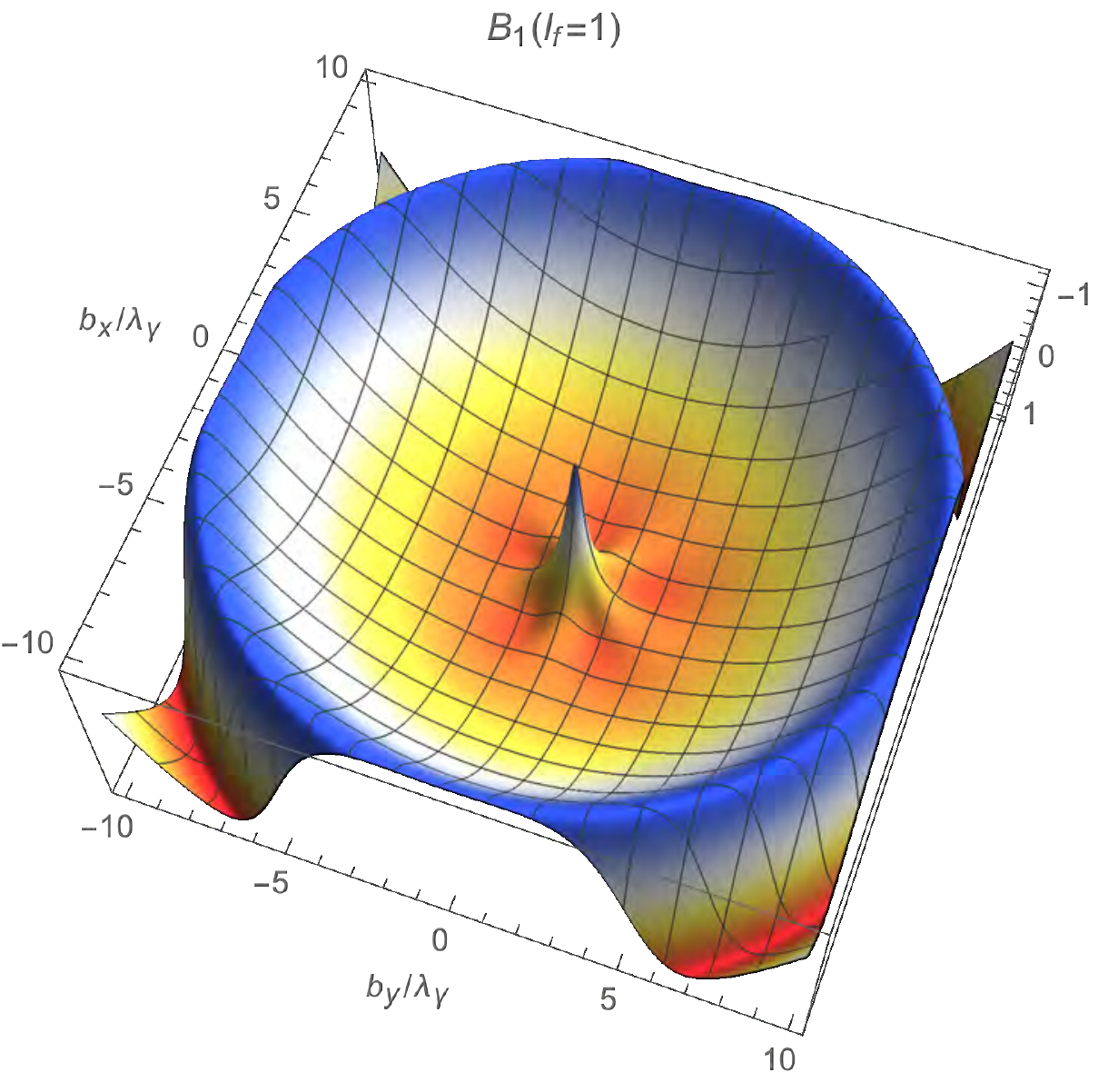}
\caption{Plots of the alignment parameters $B_2(1)$ and $B_1(1)$, top to bottom,  for $S \to P$ transitions, with contour plots on the left and 3D versions of the same on the right.   Each plot shows the $B_k$ parameter as a function of the impact parameter components $b_x$ and $b_y$ measured in wavelengths of the incident light beam.}
\label{fig:s2p}
\end{figure*}

First, we consider the $E1$ electric dipole $S\to P$ transition, for which the atom is excited into $l_f$=1 final state. Atom's alignment parameter $T_{20}(l_f=1)$ caused by the twisted photons was previously calculated in Ref.\cite{2014PhRvA..90a3425S}. The atomic polarizations $B_1(1)$ and $B_2(1)$, also known as $T_{10}(1)$ and $T_{20}(1)$, are shown in Fig.~\ref{fig:s2p}.   According to Eq.~\eqref{eq:pols2p},  they are in one-to-one correspondence (up to the overall signs) with the corresponding spin-density matrix components of the twisted photon.  The fivefold symmetry stemming from the difference in $m_\gamma$ is clearly seen in the central regions, and the graphs also show a significant spike in the $B_{k}(1)$ in the central regions.   

\begin{figure*}[h]
\centering
\includegraphics[scale = 0.43]{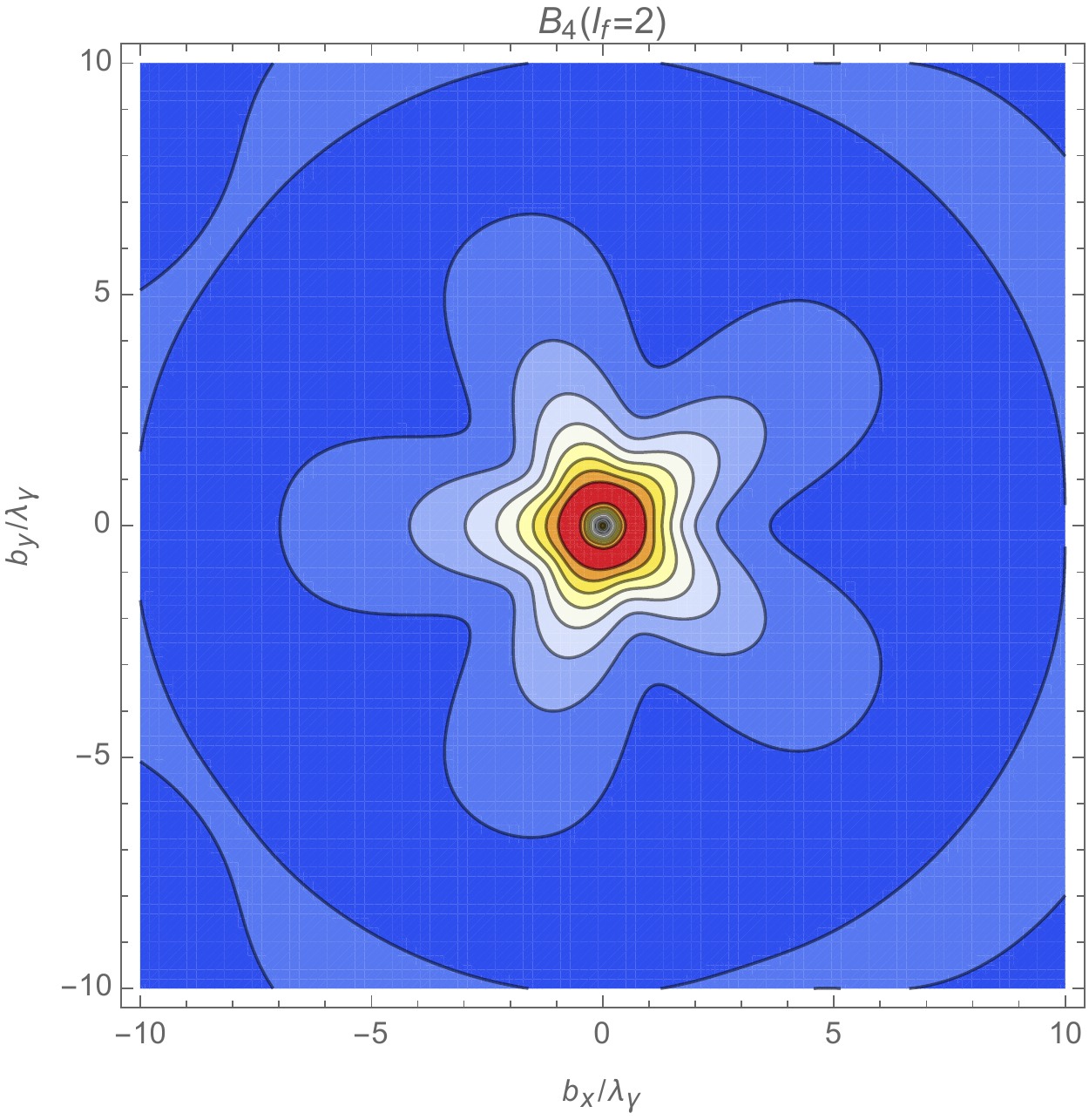} \hfil
\includegraphics[scale = 0.48]{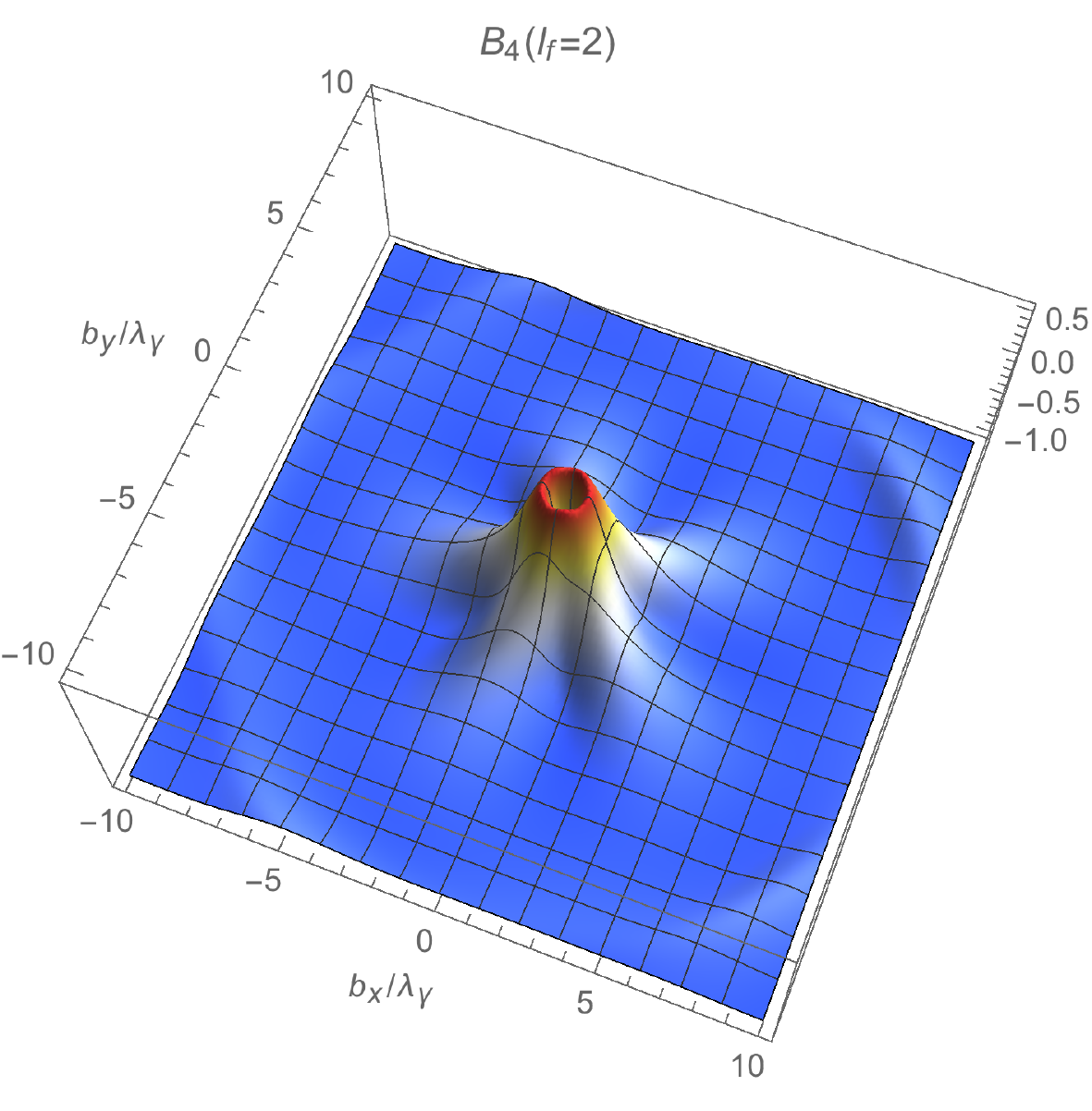}

\includegraphics[scale = 0.43]{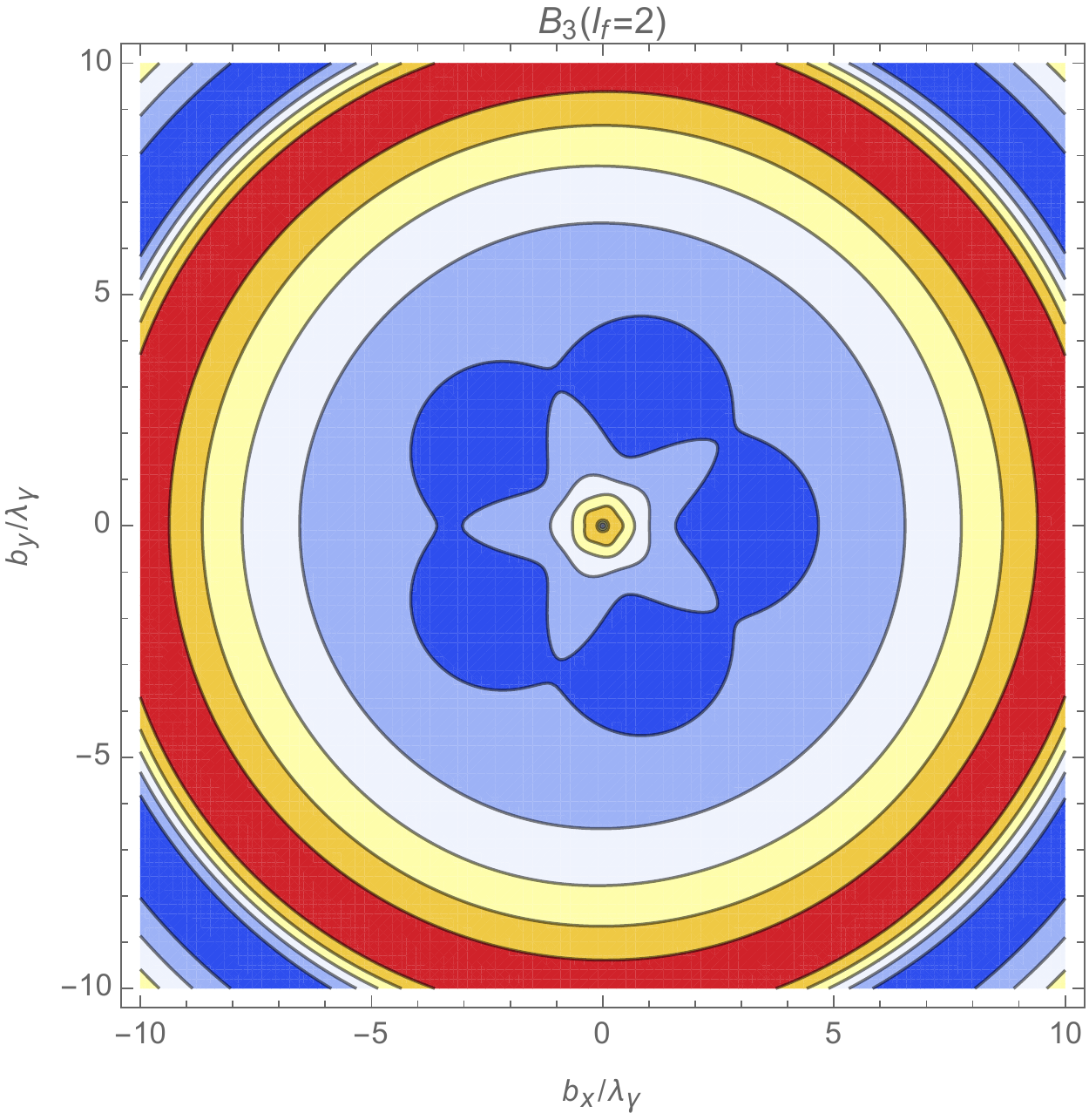} \hfil
\includegraphics[scale = 0.48]{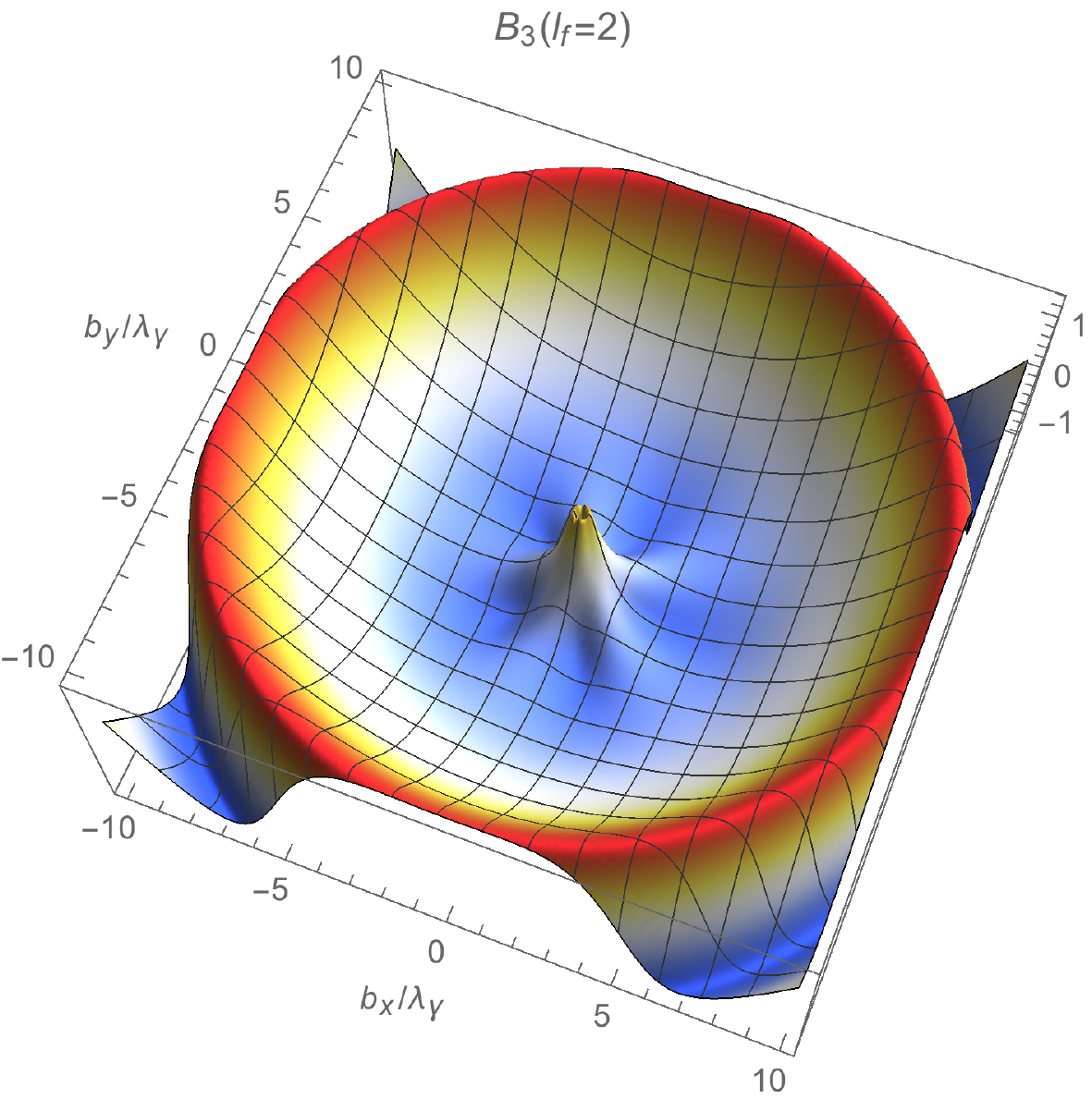}

\includegraphics[scale = 0.43]{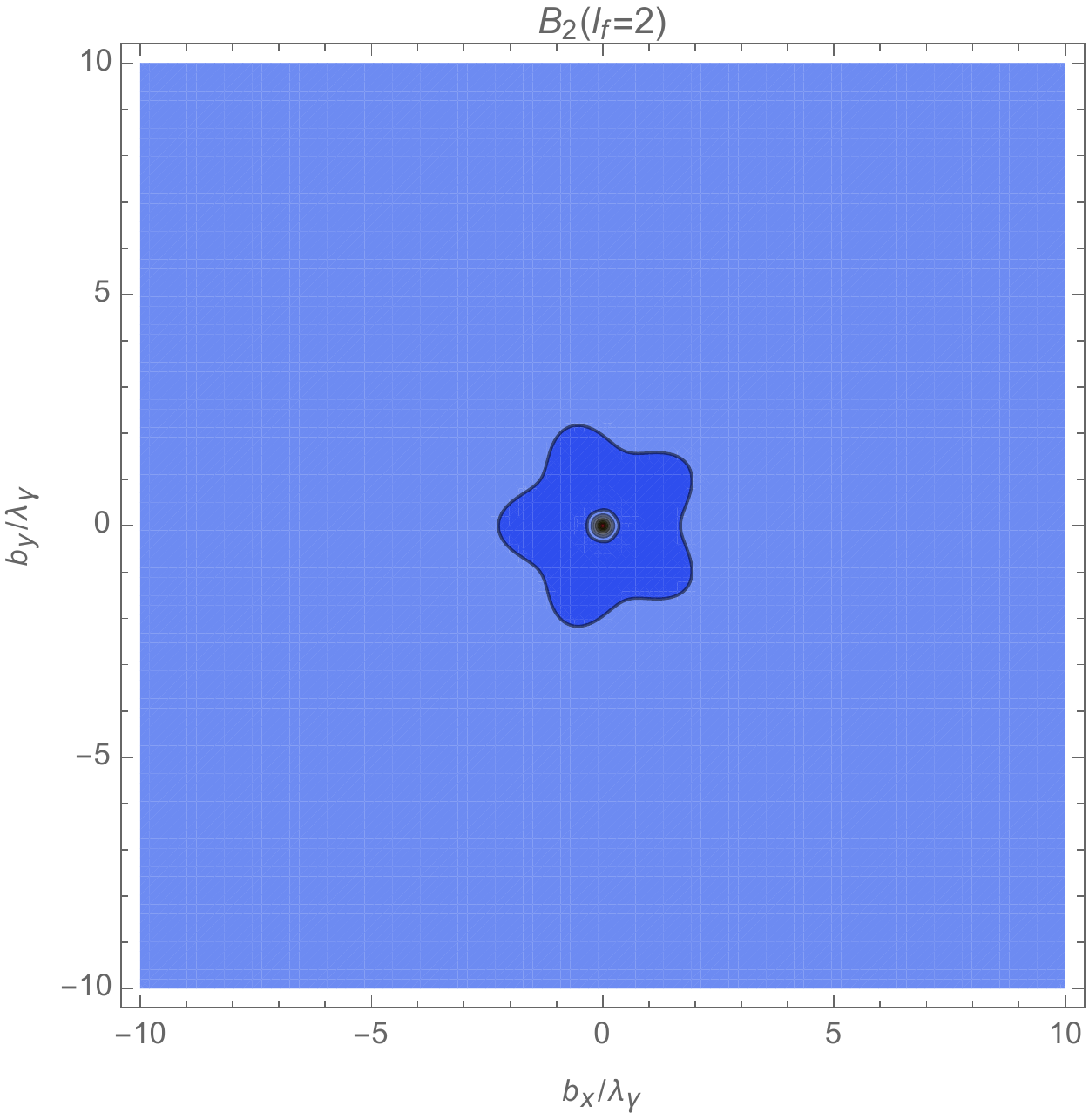} \hfil
\includegraphics[scale = 0.48]{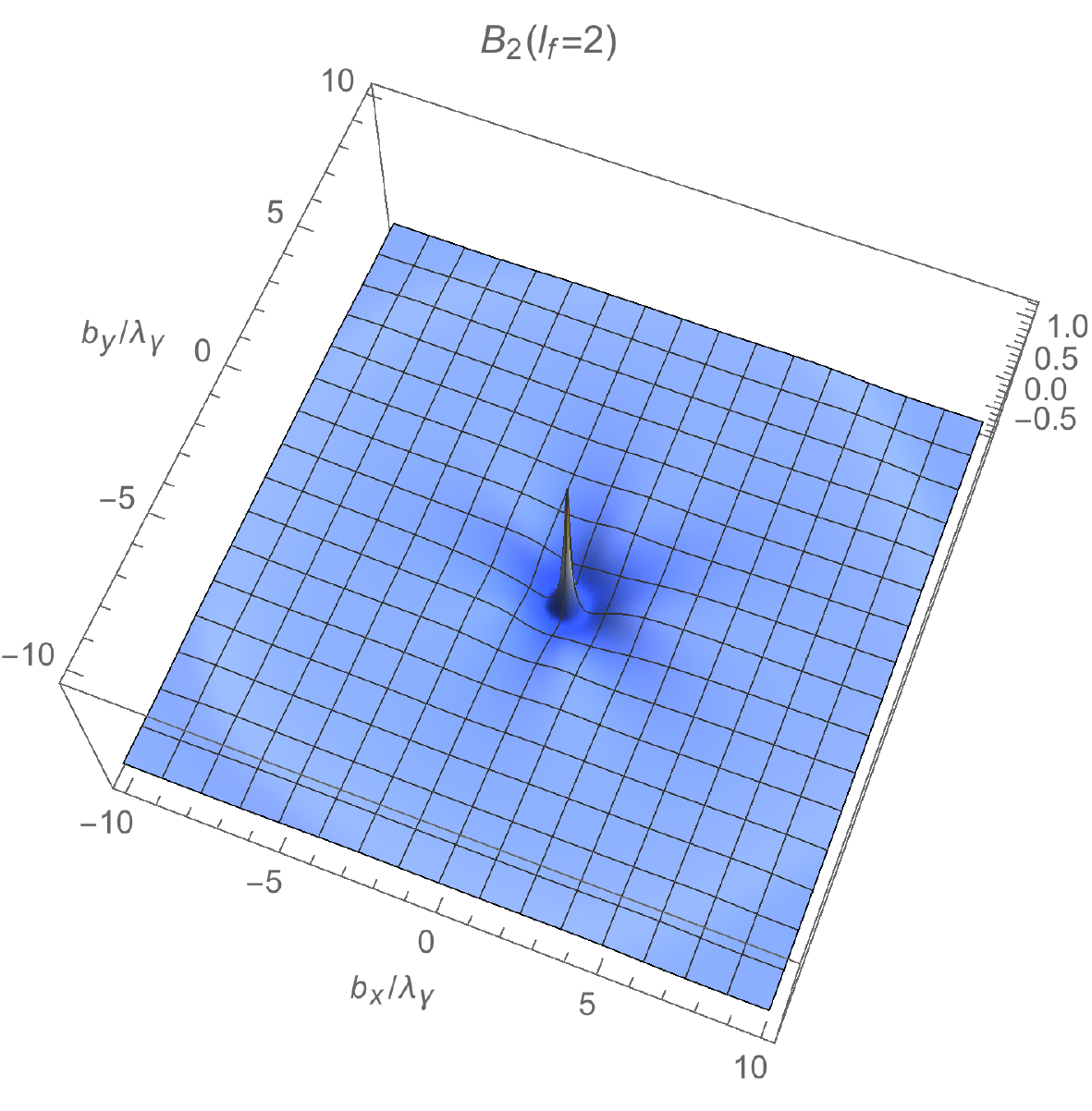}

\includegraphics[scale = 0.43]{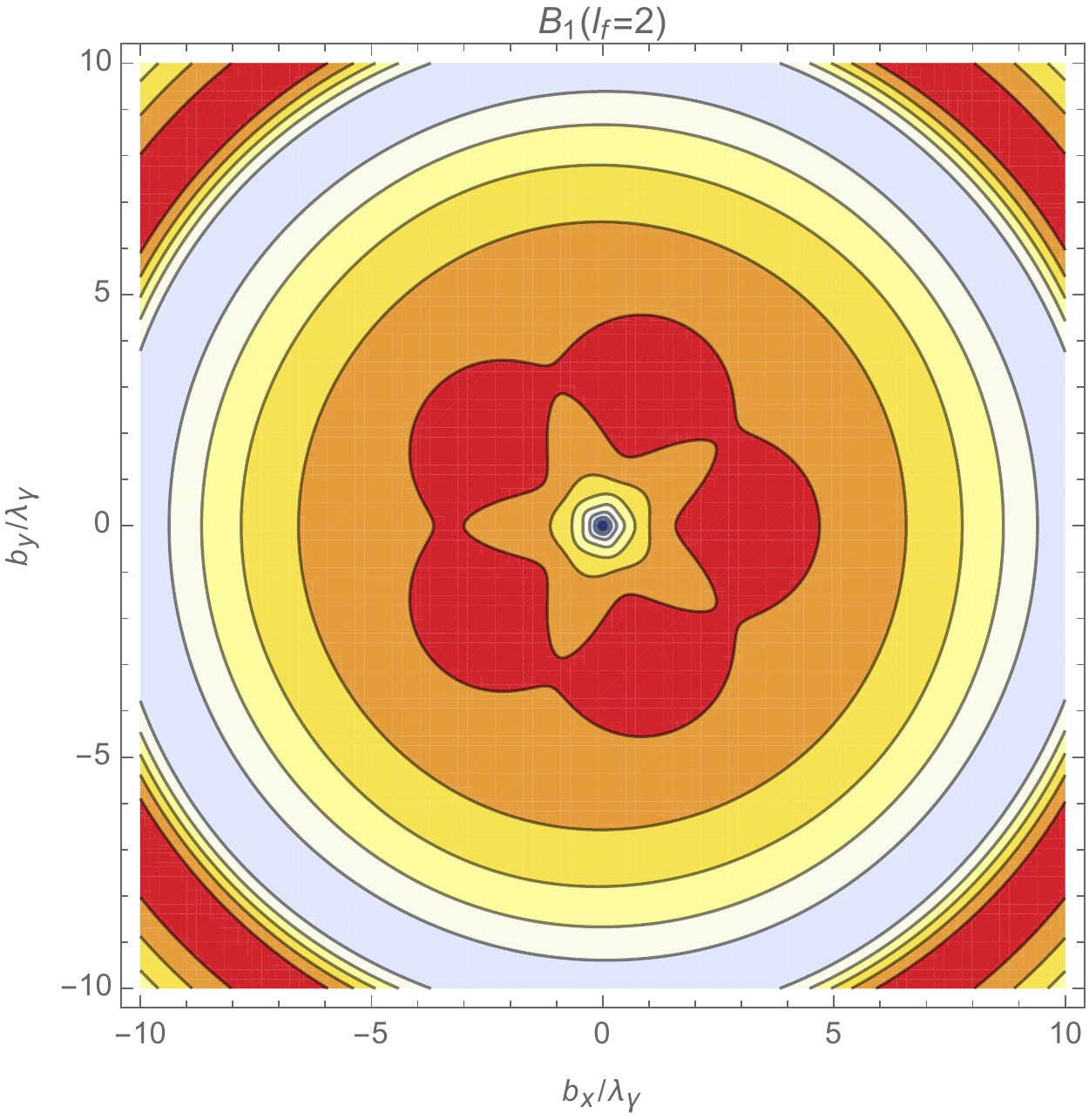} \hfil
\includegraphics[scale = 0.48]{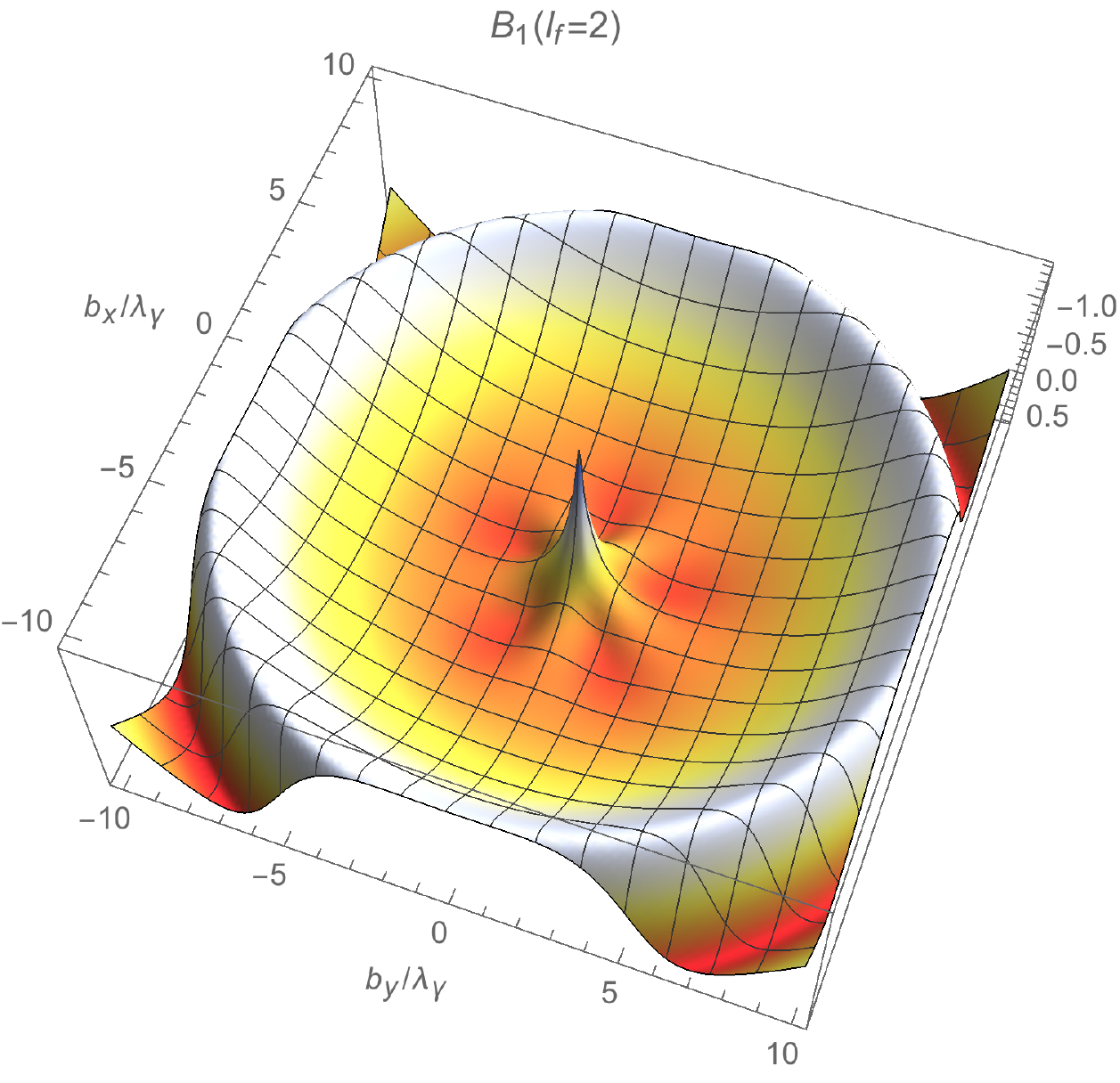}
\caption{Plots of the alignment parameters $B_4(2)$, $B_3(2)$, $B_2(2)$, and $B_1(2)$, top to bottom, for $S\to D$ transitions, with contour plots on the left and 3D versions of the same on the right.   Again, each plot shows the $B_k$ parameter as a function of the impact parameter components $b_x$ and $b_y$ measured in wavelengths of the incident light beam. }
\label{fig:s2d}
\end{figure*}

Second, we present results for atomic polarizations of $l_f$=2 final state excited in electric quadrupole $E2$-transition. In this case, there is no longer one-to-one correspondence between polarizations of the photon and the atom. The results are shown in Fig.~\ref{fig:s2d} for the orientation parameters $B_k$. Again there is a striking fivefold symmetry and central spikes of the alignment parameters are clearly visible in the central regions.  The regions farther out are more isotropic.

\section{Summary}


We have developed a theoretical formalism that describes atomic polarization due to excitation by twisted photons. Atomic photo-excitation form factors are derived with separated contributions from the longitudinal (along beam's propagation direction) and transverse electromagnetic fields. It is shown that for electric dipole $E1$ transitions the longitudinal fields contribute only to atomic transitions with unchanging magnetic quantum number $\Delta m=0$, whereas for $E2$ multipoles and higher it contributes to $\Delta m\neq 0$ excitations, as well. Parameterizing atomic polarizations in terms of a spin density matrix, we have shown that for $\ell=1$ excited atomic state, the spin density matrix is in one-two-one correspondence with a similar matrix that can be defined for a twisted photon in terms of eight independent parameters: three vector and five tensor polarizations.
For excitations of $\ell>1$-states, we use state multipoles or alignment parameters to describe atomic polarizations. It is demonstrated that the alignment parameters are sensitive to the angular momentum of the incoming field via its azimuthal asymmetries.

The presented formalism provides a framework for analysis of atomic photoexcitation with the twisted light and a tool to characterize polarization of the twisted photons with localized atomic probes available, for example, in cold-ion traps.



\section*{Acknowledgements}
The work of A.A. and H.W.~was supported by US Army Research Office Grant W911NF-19-1-0022.  C.E.C.~thanks the National Science Foundation (USA) for support under grant PHY-1812326.


\bibliography{twistedtensor}

\end{document}